# Calcium fluctuations drive morphological patterning at the onset of *Hydra* morphogenesis

Erez Braun

Department of Physics & Network Biology Research Laboratories

Technion – Israel Institute of Technology, Haifa 3200003, Israel

**ABSTRACT**

Morphogenesis in animal development involves significant morphological transitions leading to the emerging body plan of a mature animal. Understanding how the collective physical processes drive robust morphological patterning requires a coarse-grained description of the dynamics and the characterization of the underlying fields. Here I show that calcium *spatial fluctuations* serve as an integrator field of the electrical-mechanical processes of morphogenesis in whole-body *Hydra* regeneration and drive the morphological dynamics. We utilize external electric fields to control the developmental process and study a critical transition in morphogenesis, from the initial spheroidal shape of the tissue to an elongated cylindrical shape defining the body plan of a mature animal. Morphogenesis paused under external voltage is associated with a significant increase of the calcium activity compared with the activity supporting normal development. The enhanced calcium activity is characterized by intensified spatial fluctuations, extended spatial correlations across the tissue and faster temporal fluctuations. In contrast, the normal morphogenesis process is characterized by relatively moderate calcium fluctuation activity and restrained spatial correlations. Long-range communication however, is essential for development. Blocking gap-junctions halts morphogenesis by suppressing the long-range electrical communication, severely reducing the overall calcium activity and enhancing its localization in the tissue. Normal calcium activity is resumed following the wash of the blocker drug, leading to a morphological transition characterizing a normal regeneration process and the emergence of a mature animal. Our methodology of controlling morphogenesis by a physical electric field allows us to gain a global statistical view of the dynamics. It shows that the normalized calcium spatial fluctuations exhibit a universal shape distribution, across tissue samples and conditions, suggesting the existence of a global constrain over these fluctuations. Studying the correlations in space and time of the calcium fluctuation field at the onset of morphogenesis opens a new vista on this process and paints a picture of development analogous to a dynamical phase transition.

**Introduction**

Morphogenesis—the emergence of a body-plan in animal development during embryogenesis or in whole-body regeneration, is a pattern-formation process manifested by significant morphological transitions [1]. While there has been progress in deciphering the molecular and cellular mechanisms underlying morphogenesis and demonstrating their universality, a physics framework of the patterning dynamics based on organization principles and a coarse-grained description integrating the underlying processes is still missing. The common view regards biochemical processes, reaction-diffusion of morphogens, in conjunction with other signaling sources providing positional information, as the drivers of patterning in development [2, 3]. Turing's original proposal of *morphogens* as the relevant fields underlying the biochemical processes in morphogenesis opened a novel vista on developmental systems; modern variants of his model basically follow a similar concept [4, 5]. However, recent work demonstrated that mechanical [6-12] and electrical [13-18] processes play essential roles and are integrated with the biochemical processes in driving morphogenesis. Both mechanical and electrical processes allow long-range fast communications across the living tissue, beyond those supported by the slow molecular diffusion processes. Thus, mechanical stress fields, electrical voltage gradients and ionic currents are essential for closed-loop dynamics ensuring robust patterning in the face of environmental and internal fluctuations [9]. How the biochemical, mechanical and electrical processes are coordinated in morphogenesis is still a challenging open question. For the morphological patterning part, it is crucial to characterize fields integrating the electrical-mechanical processes; studying the phenomenology of such a field is a step towards a physics framework of morphogenesis.

Experimental investigations of the morphology patterning dynamics face a few challenges. Mainly, lack of controls of the morphological transitions along the developmental trajectory enabling a statistical characterization of the dynamics. Living systems are *historical* objects; each developing organism evolves idiosyncratically along its own trajectory determined by the dynamics. There is therefore inherent essential variability among samples even if they are nominally identical. Individuality and sensitivity to initial and external conditions prevent precise repetitions of experimental trajectories and thus the construction of proper statistical ensembles [19]. Developmental processes are also strictly non-stationary, challenging our attempts to analyze intrinsic observables controlling the dynamics. Development seems like a continuous "running" process advancing at its own pace. Indeed, development is commonly regarded as the result of a "program", rather than a self-organized dynamical process [19]. Without proper controls, the methodology used so successfully in studies of pattern formation in physical dynamical systems and phase transitions [20], is somewhat limited for developmental systems.

Regeneration provides a powerful model to study morphogenesis due to its flexibility, allowing to apply a wide range of experimental manipulations. This process is closely related and utilizing similar molecular biology to embryogenesis, complementing the picture of morphogenesis in animal development [9, 21]. Among models of whole-body regeneration, *Hydra*, a fresh-water multicellular organism, stands out in its remarkable regeneration capabilities; even small excised tissue segments or condensed aggregates formed from a mixture of dissociated cells, regenerate into complete animals within a couple of days [22, 23]. A tissue segment first goes through an essential stage in which it forms a hollow spheroid, made of a bilayer of epithelial cells, which eventually regenerates into the body of a mature animal [11, 24, 25]. A proper morphology can be achieved with epithelial cells alone and cell division is not an essential part of the regeneration process [23, 26, 27].

We take advantage of the flexibility of *Hydra* whole-body regeneration from a tissue segment to control the morphogenesis trajectory. Recently, we demonstrated that the application of moderate external AC electric fields allows us to modulate morphogenesis in *Hydra* regeneration on demand, halting the developmental process and even reversing it [13]. These results demonstrated the strong coupling between the electrical and mechanical processes in *Hydra* regeneration, but their integrated collective dynamics remain elusive. Direct nondestructive modulation of the course of morphogenesis provides an unprecedented tool to study the physics of the developmental dynamics in a controlled manner. As demonstrated before, $Ca^{2+}$ serves as a proxy for the underlying electrical processes in *Hydra* [13, 32, 53]. It was also demonstrated that on short time scales (seconds), $Ca^{2+}$ activity correlates with the mechanical modes of mature *Hydra* [32]. Here we utilize the electric field control to study the dynamics of the calcium ($Ca^{2+}$) field in conjunction with the tissue morphology dynamics around a critical morphological transition in *Hydra* regeneration.

*Hydra* morphogenesis is supported by its special electrical and mechanical properties. *Hydra* possess unique electrical characteristics: most cells are electrically excitable and the epithelial tissues are capable of generating and propagating electrical action potentials (even in nerve-free *Hydra*) [28-33]. Electrical excitations in the epithelial tissue can emerge spontaneously or under external stimuli. Long-range electrical communication is carried out via gap-junctions, connecting neighboring cells within each epithelial layer (ectoderm and endoderm) as well as between the two layers, allowing electrical excitations to propagate across the tissue [34-36]. From the mechanical side, the actomyosin cytoskeleton plays an essential role in *Hydra* physiology [9, 37]. The actin cytoskeleton forms supracellular networks of contractile fibers, which span the whole organism, near the basal side of each epithelial layer [38-40]. These supracellular actin fibers are organized along the animal axis in the ectoderm (the outer layer) and in a perpendicular direction in the endoderm (the inner layer) [10, 11, 41]. The fibers are connected between neighboring cells within the epithelial layers via cell-cell junctions, and are mechanically coupled to the extracellular matrix (mesoglea) connecting the two epithelial layers. In a mature *Hydra*, the contraction of the ectoderm or endoderm actomyosin fibers, respectively, enables the shrinkage or expansion of the animal body [32]. The *Hydra* epithelial tissue is therefore basically a muscle. The closest analogy to the *Hydra* tissue is that of a smooth muscle which plays important roles in different organs across organisms [42, 43]. As in other types of smooth muscles, actomyosin force generation depends on the cytoplasm free $Ca^{2+}$ concentration which can change due to influxes from the external medium or by release from internal reservoirs [44-47].

Our recent works provide evidence for the roles of the actin-fiber supracellular organization in *Hydra* regeneration, in determining the alignment of the new formed body-axis, its polarity and in providing organization centers for functional elements in morphogenesis [9-11]. *Hydra*, being a fresh-water animal, needs to balance the osmotic pressure gradients across its epithelial tissues [48]. These gradients drive water and ions through the bilayer tissue shell of the hollow spheroid in regeneration, as well as through the body-wall of a mature animal [24, 25, 49-51]. The internal fluid in the cavity enclosed by the epithelia tissues envelope, applies hydrostatic pressure forces modulated by the inflow of water driven by the osmotic pressure gradients. Taken together, two types of mechanical forces provide major contributions to the morphology of the *Hydra* tissue in regeneration; the hydrostatic fluid forces, modulated by the osmotic pressure gradients, that are approximately isotropic and tend to inflate the enclosed tissue [52], and the internal contractile forces generated by the active actomyosin supracellular fibers [11].

The *Hydra* tissue undergoes a clear morphological transition along the regeneration trajectory, from an initial spheroidal shape to that of a cylindrical shape characterizing the body plan of a mature animal. This critical morphological transition is a dramatic and significant change driven by the mechanical processes along the tissue's developmental trajectory towards regeneration. It is not clear how sharp this transition is and whether it involves threshold-crossing of certain processes or symmetry-breaking events. The external electric field control of the course of morphogenesis allows us to characterize the $Ca^{2+}$ spatio-temporal dynamics in conjunction with the morphology patterning, driven by the underlying mechanical processes in the tissue, at the onset of this critical transition for extended periods by pausing the developmental process on demand. The analysis of the $Ca^{2+}$ field is challenging due to deformations of the tissue in 3D which, unlike e.g., in many neural systems, do not allow association of the signals with definite locations in the tissue over time. To overcome this limitation, we take a statistical approach and attempt to characterize the statistics of the $Ca^{2+}$ field, rather than following its dynamics at assigned tissue locations over time.

Our analysis shows that the $Ca^{2+}$ *spatial fluctuations* are strongly correlated with the morphology dynamics, suggesting these fluctuations as an underlying integrator field of the electrical-mechanical processes in morphogenesis. Switching-on the external voltage control causes a qualitative change in the $Ca^{2+}$ spatio-temporal dynamics. This change in the $Ca^{2+}$ dynamics is associated with the pause of the morphogenesis process, manifested in the inhibition of the critical morphological transition normally observed along the regeneration process. Pausing regeneration is reversible and normal morphogenesis is resumed following the switching-off of the external voltage [13]. It allows us to identify the essential statistical characteristics of the $Ca^{2+}$ fluctuations in space and time supporting morphogenesis. In particular it shows that an enhanced calcium activity, characterized by intensified spatial fluctuations, extended spatial correlations across the tissue and fast temporal fluctuations is associated with halting the proceeding of morphogenesis along its natural trajectory. The strong associations of the morphology patterning with the $Ca^{2+}$ spatial fluctuations, raises a question about the role of the tissue's long-range electrical communication in morphogenesis. We show that blocking gap-junctions in the regenerating *Hydra* tissue prevents morphology patterning in a reversible way, demonstrating the importance of an extended communication in morphogenesis. The surprising ability to control regeneration on demand, either by an external voltage or by blocking gap-junctions, without damaging the regeneration potential of the tissue sheds new light on the physics of morphogenesis. Identifying the $Ca^{2+}$ spatial fluctuations as a relevant field underlying a morphological transition, suggests an analogy with bifurcations in physical pattern formation processes or dynamical phase transitions.

The structure of this paper is as follows: First, we take a global view of the dynamics by measuring different observables determining the tissue's morphology and $Ca^{2+}$ signals and follow them before and after the external voltage switch. We show that the switch-on of the voltage indeed pauses morphogenesis at the onset of a morphological transition. Next, we identify the $Ca^{2+}$ spatial fluctuations as the field associated with the morphology dynamics and quantify their correlations. We then concentrate on the $Ca^{2+}$ spatial distributions across the tissue; the localization of this field and its spatial correlation functions. Following this, we demonstrate that long-range communication supporting the calcium patterning across the tissue is essential for morphogenesis. Finally, we characterize the global statistics of the $Ca^{2+}$ spatial fluctuations under the different experimental conditions and show their universality.

**Results and Discussion**

Tissue fragments are excised from the gastric (middle) regions of mature *Hydra*, allowed to fold into spheroids for ~3 hrs and then placed in the experimental setup under a fluorescence microscope. We utilize a transgenic *Hydra* expressing a fast $Ca^{2+}$ fluorescence probe in its epithelial cells, allowing to follow the calcium activity over the duration of the experiments [13] (Methods).

We first seek to develop a global view of the dynamics by characterizing the major morphological and $Ca^{2+}$ observables under the external voltage control. The voltage protocol for this experiment is depicted schematically in Fig. 1a. The traces in Fig. 1b show the dynamics measured by time-lapse microscopy at 1 min resolution, first at zero voltage followed by switching the external voltage to V=40V (Methods). These traces are typical examples of the dynamics observed in three repeated experiments, with similar analysis carried over for more than 11 different tissue samples in these experiments. More examples are shown in the supplementary material. The morphology is characterized by the tissue's projected *area* and *aspect-ratio* (AR; short axis/long axis of a corresponding ellipsoid), which serve as the first two moments characterizing its shape modes. The $Ca^{2+}$ signal is characterized by the tissue's fluorescence spatial *mean* (total fluorescence/area) and its spatial *variance* (Vr), indicating the spatial fluctuations of the field, which serve as the first two moments of the $Ca^{2+}$ activity. The total fluorescence, measuring the overall $Ca^{2+}$ activity over the tissue independent of the projected area, is also presented for comparison. Finally, the fluorescence signal spatial *standard-deviation over mean* (Std/mean) measures the normalized $Ca^{2+}$ spatial fluctuations in the tissue, allowing an unbiased comparison between different tissues and conditions. See Methods for the precise definitions of these observables. The 1 min time resolution is chosen to allow on the one hand, faithful measurements of the dynamics for multiple samples and on the other hand to follow a few cycles of regeneration over days, without apparent damage to the tissues. Higher temporal resolution measurements might reveal more information on fast kinetics [32], but do not seem to change the picture on the long-term dynamics of the $Ca^{2+}$ fluctuations, of interest to this work.

At zero voltage the tissue exhibits a relatively sparse $Ca^{2+}$ activity in the form of sharp spikes, accompanied by somewhat regular area and AR dynamics. $Ca^{2+}$ spikes at zero voltage reflect spontaneous excitations of the unstimulated tissue. The tissue's projected area exhibits a somewhat regular sawtooth-like dynamics under this condition. These pattern of the area dynamics is commonly observed for an undisturbed regenerating tissue and is thought to be driven by the osmotic pressure gradients, leading to a relatively smooth inflation of the spheroid tissue due to water influx followed by a sudden collapse of the tissue due its local rupture releasing the pressure [11, 24, 25, 50]. The constant-rate increases in area during the inflation periods are often associated with exponential decays, decorated with large fluctuations, of the $Ca^{2+}$ Vr signals (note the Vr y-axis log scale). The observed slow, relatively smooth, change in the AR along the zero voltage trace, is due to an elongation episode of the tissue followed by its relaxation, as part of its normal dynamics towards regeneration.

Switching on the external voltage (red line in Fig. 1b), shows a considerable qualitative change in the dynamics; all observables now exhibit highly fluctuating behavior while the tissue still does not regenerate at the end of the displayed time trace. Fig.1c shows representative examples of bright-field (BF) and fluorescence images corresponding to different stages of the above dynamics (see also Supplemental Movies 1,2). The different measures of the $Ca^{2+}$ activity (total, mean, Vr and Std/mean fluorescence signals) show some similarities in their traces under both conditions, with and without the application of the external voltage, but with important differences in their detailed dynamics; the Vr and Std/mean observables show richer dynamics than the total or mean fluorescence (see Fig. S1 for detailed comparison of the different observables). While the total or mean fluorescence measure only global signals, the Vr detects the degree of non-uniformity of the spatial distribution of the $Ca^{2+}$ activity across the tissue at every time point and provides more information on the activity (see the example images in Fig. 1c). The analysis below quantifies these spatial dynamics. In the following, we concentrate on the tissue projected area and AR as proxies of the tissue's morphology, and on the spatial Vr (second moment) and the normalized spatial fluctuations (Std/mean) as the relevant observables of the $Ca^{2+}$ fluorescence dynamics. Our concentration on the spatial fluctuations rather than the total or mean $Ca^{2+}$ fluorescence signal allows us to expose finer details of the dynamics and connect the $Ca^{2+}$ spatial activity with its temporal behavior. We later discuss the additional information provided by the spatial normalized fluctuations, the Std/mean, and their unique universal statistical characteristics.

As shown before, the effect of the external voltage on *Hydra* regeneration is reversible, allowing repeated cycles of backward and forward transitions of morphogenesis [13]. Following the voltage episode at the end of the trace in Fig. 1b, switching the voltage back to zero leads to fast regeneration, marked by persistent elongations of the tissue with eventually a transition from a spheroidal shape to a persistent cylindrical shape and the appearance of tentacles marking the completion of the regeneration process (Figs. 1d,e; the traces are a direct continuation of the ones in Fig. 1b; see Supplemental Movie 3). Such persistent elongations of the tissue are not observed in our experiments for tissues under an external voltage above threshold [13]. Switching the voltage on and off therefore serves as a control of this critical morphological transition. As shown by us before, the alignment of the body-axis in regeneration is determined from early stages by the ectoderm supracellular actin fibers, which therefore also determine the direction of the tissue's persistent elongations since the mature *Hydra* eventually develops along this axis [10, 11]. This suggests that indeed the forces generated by the long-range actin-fiber organization, enabled by the $Ca^{2+}$ signals, play an important role in the morphological transition to a persistent cylindrical shape. The extremely short regeneration time (indicated by the appearance of tentacles) after switching-off the voltage, less than 200 min, suggests that the external voltage halts regeneration at the onset of morphogenesis without damaging it. For different tissues and in separate experiments, we measure an average regeneration time of 350 min after resetting a voltage episode to zero (with a similar voltage protocol as shown here; 7 different tissue samples, with a longest time of 480 min), compared to the measured normal regeneration time of undisturbed tissue samples of 900-2900 min [11]. The precise regeneration time after switching the voltage off depends on the tissue's individuality, its history and the details of the applied voltage protocol along the experiment. Halting regeneration at the onset of morphogenesis is also manifested in the observed tissue dynamics exhibiting sporadic unsuccessful attempts to elongate significantly and persistently under the applied voltage, as is normally observed for an undisturbed tissue along its regeneration trajectory. Following regeneration, we switch the voltage on again and follow the dynamics of the mature animal. The traces in Fig. 1d show the change in dynamics following a second round of high voltage (V=50V) starting at the red line. Reversal of the morphology from an elongated body-plan of the mature animal is manifested in the measured area and AR of the tissue sample, which folds the tentacles and reverts to a spheroidal shape similar to the morphology before regeneration (see the corresponding images in Fig. 1e and Supplemental Movie 4). Under the voltage, the tissue sample again does not exhibit episodes of persistent elongations. The morphological transition back to a spheroidal shape is accompanied by the resumption of strong fluctuations of the morphological observables as well as of the spatial variance (Vr) of the $Ca^{2+}$ fluorescence signal. This behavior gives further support for our qualitative observation of the associations between the $Ca^{2+}$ spatial fluctuation dynamics and the critical morphological transition to a persistent cylindrical shape in regeneration. More example traces are shown in Fig. S2. These experiments demonstrate that by the application of external voltage, we can control morphogenesis and gain statistical information on the $Ca^{2+}$ fluctuations and their correlations with morphological changes at the *onset of a morphological transition* in regeneration over extended periods.

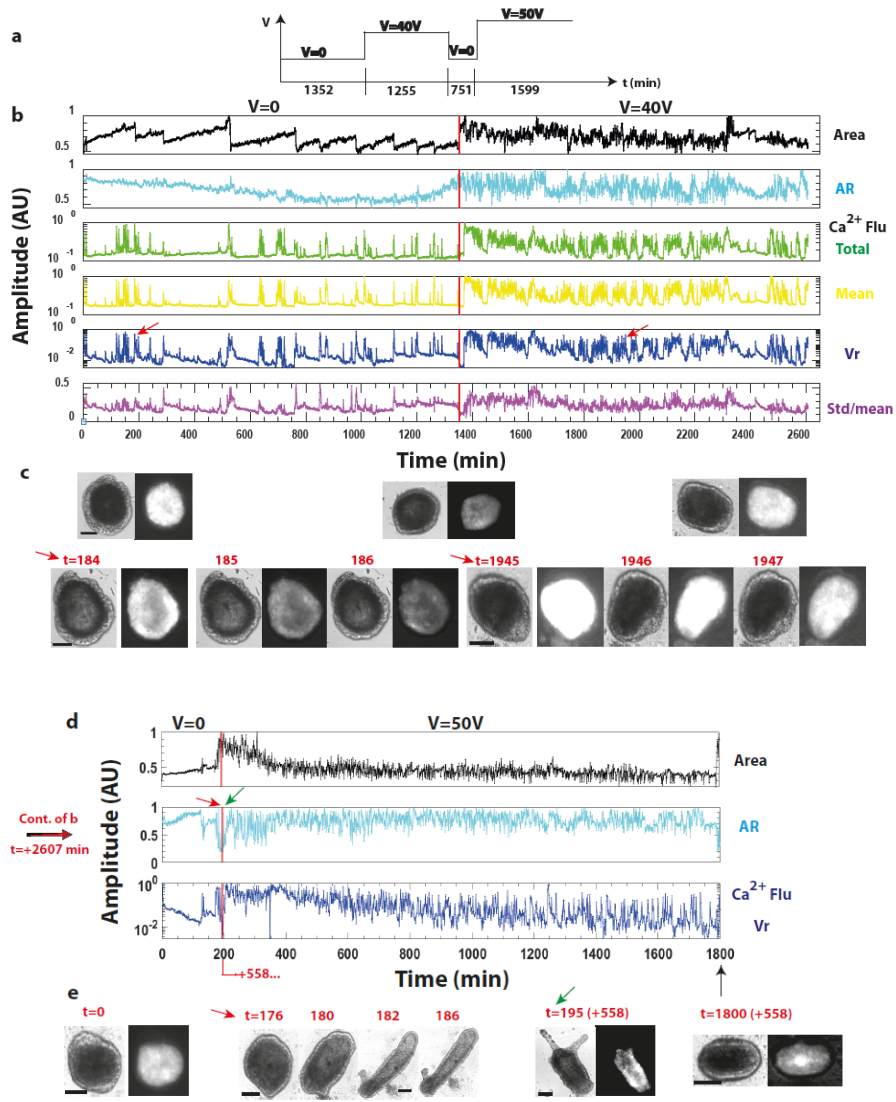

**Fig. 1: Morphology and Ca$^{2+}$ fluctuation dynamics.** (a) Schematics of the applied external voltage level along the experiment. In practice, the voltage increase from 0 to 40V and from 0 to 50V, is gradual (but over a short time compared to the trace) to prevent damage to the tissue. (b) Time traces of a spheroid made out of a tissue segment excised from a mature transgenic *Hydra* expressing a fast Ca$^{2+}$ fluorescence probe in its epithelial cells and measured by a time-lapse microscope at 1 min resolution. Traces from the top; morphological characteristics: projected *area* of the tissue (normalized by the max over the trace) and its *aspect-ratio* (AR; short axis/long axis of a corresponding ellipsoid); Ca$^{2+}$ fluorescence: *total, mean* and spatial *variance* (Vr) of the florescence signal across the tissue (all signals normalized by their max values over the trace; note the y-axis log scale for these signals), and the spatial *standard-deviation over mean* (Std/mean) of the fluorescence signal, measured at each time point. The traces to the left of the red line are at zero voltage while those to the right follow the switch of the voltage to 40V. (c) Top: Pairs of bright-field (BF) and fluorescence images at the beginning of the time traces at zero voltage (left), at the transition marked by the red line (middle) and at the end of the traces under voltage (right). Bottom: Pairs of BF and fluorescence images around example peaks of the Vr, marked by red arrows on its trace. (d) Time traces of the same tissue directly continuing the traces in Fig.1a (at t=2607 min), following switching-off the external voltage at the end of the V=40V traces and switching it again to V=50V (at the red line). Reversal of morphogenesis requires higher voltages at each round [13]. Upper panel: the tissue projected *area* (normalized by the max over the trace)*,* middle panel: *AR* and lower panel: the spatial *Vr*

(normalized by the max over the trace; note the y-axis log scale) of the $Ca^{2+}$ fluorescence. Note that measurements of the zero voltage trace actually extend to 751 min; for clarity we cut and connect the two traces in Fig. 1d at t=193 min, since after that time the tissue is fully regenerated (reflected in an increase of the tissue area and large change in AR) and the strong movements of the mature animal gets it occasionally out of focus, making the measurements unreliable. (e) Pairs of BF and fluorescence images of the tissue at the beginning of the traces (left) and a series of BF images showing fast regeneration of the tissue (red arrow), and the fully regenerated animal at the end of the V=0 trace (green arrow; adding 558 min to the time shown on the trace); note the transition of morphology from the incipient spheroidal shape to a persistent cylindrical shape. Following the voltage switch-on again to 50V, a pair of BF and flu images towards the end of the traces, showing a morphological transition back to a spheroidal shape (right; black arrow). The time markings above the images are in minutes. Bars: 100μm.

We now show that the morphology dynamics are correlated with the $Ca^{2+}$ spatial fluctuations as measured by the fluorescence spatial fluctuation observables, the Vr and the Std/mean. A direct inspection of the time traces in Fig. 2a, demonstrates that the largest peaks of the Vr signal are associated with large morphological changes. The red lines in Fig. 2a guide the eye to the correspondence between the large Vr peaks and the time points along the area trace, for zero voltage and for V=40V, for the smoothed traces of Fig. 1a (smoothing is applied here to amplify the large peaks of activity; see Methods). Under both conditions, changes in the Vr are associated with *either increase or decrease* of the tissue area. The associations between the Vr and the tissue area are directly observed even for small morphological changes; a closer inspection of Fig. 2a in fact shows that every change in morphology is associated with a change in the level of the $Ca^{2+}$ fluorescence spatial variance as measured by Vr. This can also be directly observed by comparing the mutual dynamics of the bright-field and the fluorescence images (see Supplemental Movies1,2). The opposite however is not necessarily true; changes in the Vr might emerge without apparent effects on the tissue's morphology (using the projected area as a crude proxy). Importantly, the level of the Vr signal usually starts to *increase before* noticeable area changes are detected (see Fig. S3 for a zoom view over several peaks, corresponding images and another example trace). This indicates the non-trivial relations between the tissue morphology and the $Ca^{2+}$ spatial fluctuations and suggests that the calcium activity and its patterning is an important player in driving morphological changes.

We quantify the associations between the $Ca^{2+}$ dynamics and the tissue morphology by their cross-correlation functions (Fig. 2b). The tissue's area trace is clearly non-stationary and at zero voltage is dominated by the long sawtooth structure. Our analysis of the cross-correlation functions therefore concentrates around the zero time-lag region, which is less sensitive to the non-stationary nature of the signal; the interpretation of the behavior at large time-lags is more challenging. The area-Vr and area-Std/mean cross-correlations, computed for the (unsmoothed) data of Fig. 1a, exhibit on average a negative peak for zero voltage (left) and a positive peak for V=40V (right), at a zero time-lag (sharper and stronger peaks for the Std/mean). These cross-correlations are typical in our experiments, but can also show different features for different tissues; all tissue samples show significant cross-correlations between the area and the $Ca^{2+}$ spatial fluctuations (Fig. S4). The correspondence between the tissue's area and the $Ca^{2+}$ spatial fluctuations at zero time-lag is also directly apparent in the scatter plots shown in the inset of Fig. 2b. It demonstrates the strong associations between the $Ca^{2+}$ fluorescence spatial fluctuations and the shape modes of the tissue. The relaxations of the cross-correlation functions suggest an underlying time scale of roughly 100 min in the dynamics. The autocorrelation functions for the area and Std/mean (Fig. 2c) show a similar relaxation time of roughly 100 min for both conditions, with and without the external voltage. This implies that this relaxation time is intrinsic to the tissue dynamics and does not reflect the more special sawtooth structure of the area at zero voltage. Indeed, in some cases under voltage, the tissue exhibits remarkable oscillatory dynamics with a similar time scale of roughly 100 min (Figs. 2c-right); a time scale repeatedly observed across experiments (see more examples in Figs. S4). The source of this time scale is not clear; it extends well beyond the time scales of the molecular kinetic processes, but might be compatible with long-range diffusion, the slow evacuation of the cytoplasm $Ca^{2+}$, or electrical processes coupled with a mechanical response.

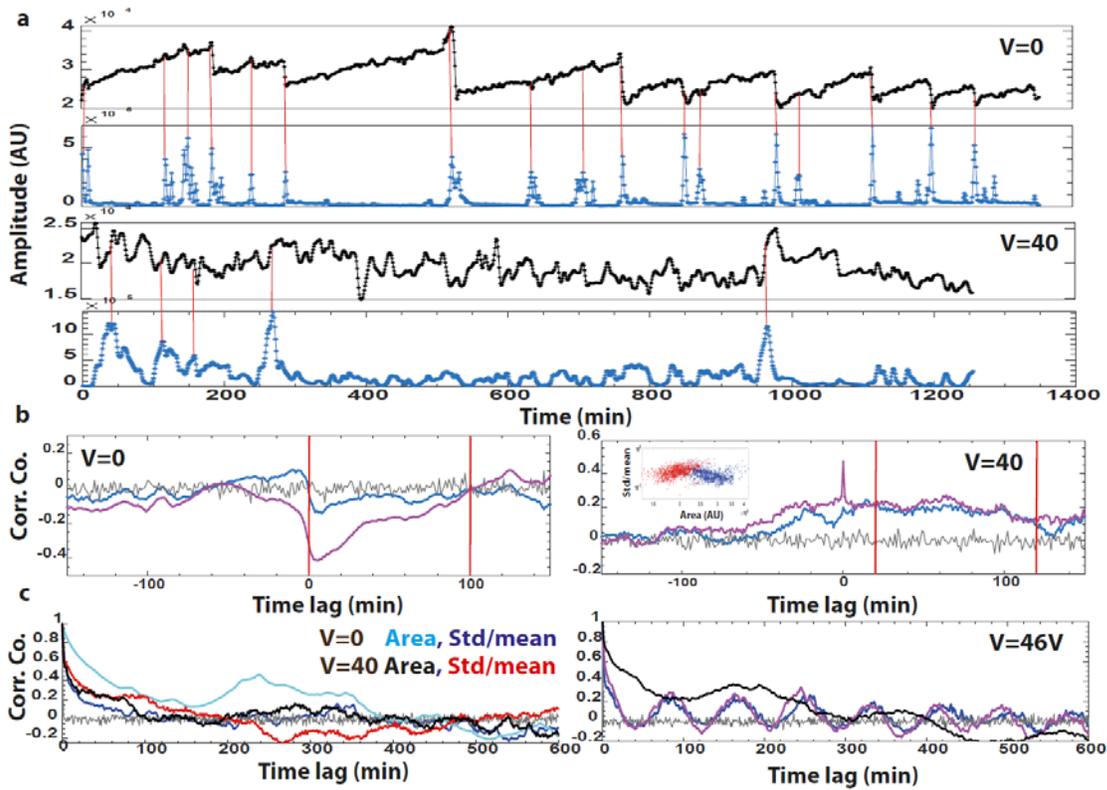

**Fig. 2: Morphology-Ca$^{2+}$ correlations.** (a-top) Smoothed time traces of the tissue projected area (black) and the fluorescence spatial Vr (blue) for zero voltage and (a-bottom) V=40V for the same traces as in Fig. 1a. Smoothing is done by a Gaussian-weighted moving average over a window of 10 min to highlight the largest peaks of activity (Methods). Note that in contrast to Fig. 1a, the Vr traces are plotted here with the y-axes not on log scale. The red lines connect the largest Vr peaks with the corresponding time points along the area trace to guide the eye. Note that the Vr signal typically starts to increase before a corresponding change in the area is detected (each dot is a measurement point at 1 min resolution). (b) Area-Vr (blue) and area-Std/mean (magenta) cross-correlation functions for the same (unsmoothed) traces as in Fig. 1a for (left) V=0 and (right) V=40V (Methods). Note the negative cross-correlations for V=0 and positive cross-correlations for V=40V at a zero time-lag. Note also the relaxation time scale of roughly 100 min for both conditions (marked by the separation of the two red lines). The inset at right shows scatter plots of the Std/mean versus area for V=0 (blue) and V=40V (red) for all time points of the trace, highlighting the negative and positive correlations respectively at a zero time-lag. (c-left) Area and Std/mean autocorrelation functions for V=0 and V=40V computed for the same traces as in (b). (c-right) The autocorrelation function (area-black, Vr-blue and Std/mean-magenta) showing an example of oscillatory dynamics of the Ca$^{2+}$ activity for a tissue under V=46V. The tissue went through two cycles of voltage switches: from zero voltage to 40V, back to zero and then switching the voltage to 46V. The gray curves in all autocorrelations and cross-correlations plots is computed from surrogate data, indicating the level of noise in the computations (Methods).

The statistical measures of the $Ca^{2+}$ spatial fluctuations (Vr and Std/mean) do not characterize the calcium spatial distributions across the tissue. We expect the spatial patterns of the $Ca^{2+}$ signal to affect the tissue's morphology, since the local calcium signal is able to modulate the spatial distribution of the mechanical forces. The *Hydra* epithelial cells are electrically coupled via gap-junctions. In mature animals, electrically excitable systems with such strong couplings usually exhibit large-scale spatial synchronization of the signal, but this was not explored in regenerating *Hydra* tissues. Therefore, to gain insight on the patterning dynamics, we next seek to analyze the spatial distribution of the $Ca^{2+}$ activity over the tissue and the effect of the voltage switch on the spatial correlations of this activity. A simplified qualitative glance over the spatial distributions of the $ca^{2+}$ activity over time is presented by highlighting, at every time point, the relative locations of its strongest signals in the tissue. To reduce the noise level, we first compute the coarse-grained $Ca^{2+}$ fluorescence signals by averaging them within triangular mesh elements, covering the tissue at each time point (see Methods for details of the mesh construction [54] and the coarse-graining procedure). Fig. 3a shows a time trace of the locations of the strongest coarse-grained $Ca^{2+}$ activities overlaid with the tissue's area trace (black curve), at zero voltage and after switching the voltage to V=40V at the red line, for the same experiment as in Fig. 1a. At every time point we divide the entire tissue into quarters and plot the distance r of the centroids of mesh elements from the tissue's centroid, showing a coarse-grained $Ca^{2+}$ signal above an *arbitrary threshold*. Different colors mark different quarters (shown schematically in Fig. 3a). Detailed explanation of the definition of the tissue's quarters and the identification of the tissue's centroid can be found in the Methods. The distance r is normalized by the maximal tissue's radius at each time point. We keep the same definition of quarters in the lab frame for the entire trace. However, since the tissue can freely rotate the marked quarters are not fixed with respect to the tissue, but within a reasonable time period can actually remain the same. At a given time point, the dispersion in r and the distribution among quarters provide a qualitative view of the level of localization of the strongest $Ca^{2+}$ signals. Fig. 3a shows that the $Ca^{2+}$ activity is significantly more localized at zero voltage than at high voltage. This relative localization is characterized by isolated islands of activity that do not widely spread over r and a small color variation at a given time point, reflecting a narrow distribution of the signal among the different quarters. The precise value of the threshold does not affect this qualitative picture.

We next quantify the spatial correlations. The normalized covariance matrix of the $Ca^{2+}$ fluorescence fluctuations is computed at each time point, using normalized signals; subtracting from the coarse-grained amplitudes the spatial mean and dividing by the spatial standard-deviation of the fluorescence levels over the tissue at this time point. We divide the maximal distance ($R_{max}$) between mesh triangles at each time point into several bins and average the normalized covariance for all pairs residing within a certain distance range to get an estimate of the correlation coefficient at this length scale. The trace in Fig. 3b shows the correlation coefficients at $0.08<R<0.12\ R_{max}$ (green) as a function of time for the same experiment as in Fig. 3a. Apparently, the spatial correlation coefficients strongly fluctuates in time under all conditions, and their dynamics are somewhat associated with the spatial Vr, but do not exactly follow it. Detailed examination demonstrates that this correspondence is due to associations of both the spatial correlations and the spatial fluctuations with the overall $Ca^{2+}$ activity (see Figs. S5). We further elaborate on this important observation below. The time-averaged correlation function decays on shorter scales for zero voltage compared to the longer-range decay at V=40V, and is reduced to zero at ~0.25 $R_{max}$ for the latter condition (Fig. 3c). To gain a statistical view of the spatial correlations across tissue samples, we treat the time trace as a statistical ensemble and compute the distributions of the spatial correlation coefficients for four length-scale ranges across the tissue (Fig. 3d). These distributions accumulate statistics from 10 sample tissues in 3 different experiments (>11,000 measurement points for each distribution). The tissues under external voltage show significantly more extended correlations compared with those under zero voltage. The large spatial fluctuations of the $Ca^{2+}$ activity (manifested in the Vr or Std/mean) together with its rather limited spatial correlations, which even under the external voltage do not significantly extend beyond $R>0.2R_{max}$, demonstrate that the tissue is not uniformly excited even under the constant external voltage stimulation. Rather, excitations are usually manifested by the emergence of limited-range activity islands, which eventually decay sporadically rather than in synchrony (see example images in Figs. S3, S5). The associations of the overall activity with both the level of spatial fluctuations and with the level of the spatial correlations, indicate the unusual characteristics of the tissue's excitation modes; these associations presumably play important roles in supporting morphogenesis. The weaker spatial correlations of the $Ca^{2+}$ signal for zero voltage compared to those under voltage, together with the stronger localization of the signal under this condition, suggest that regeneration favors moderate synchronization of the $Ca^{2+}$ activity across the tissue; stronger synchronization and stronger calcium activity inhibit morphological patterning and are unfavorable for proper morphogenesis. More examples supporting this observation are shown in Figs. S2, S5.

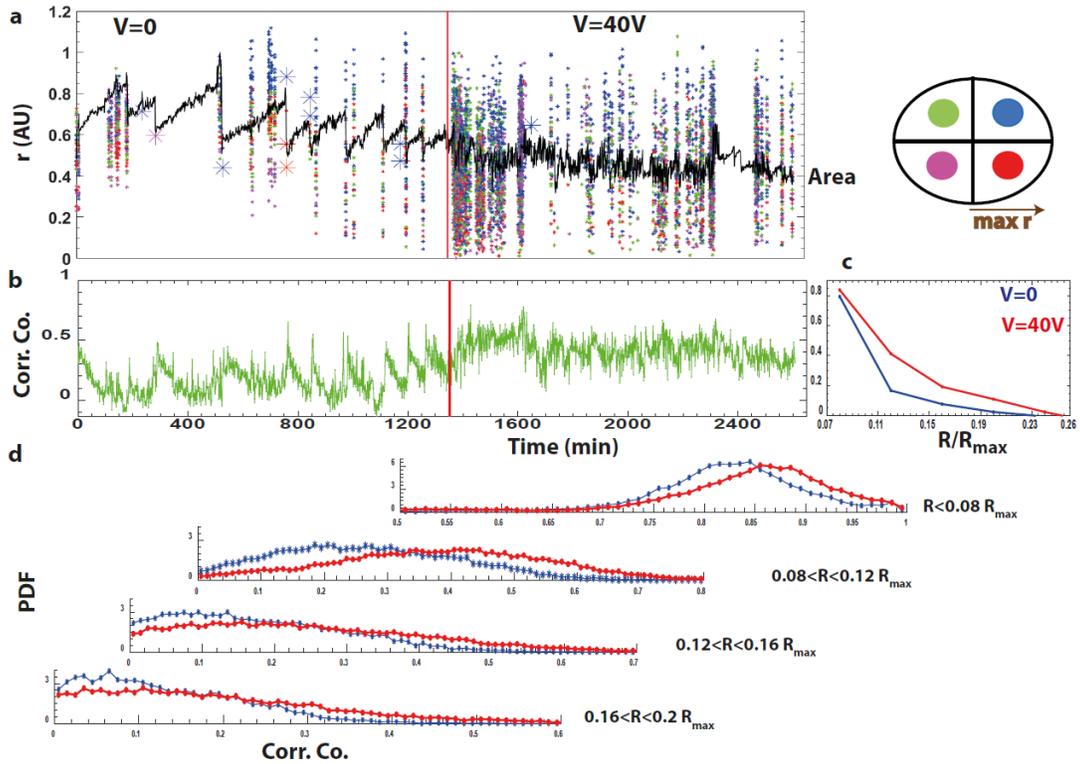

**Fig. 3: Ca$^{2+}$ localization and spatial correlations.** (a) Time traces of the projected *area* (black curve; normalized by the max over the trace) and r, the location of the coarse-grained Ca$^{2+}$ signals above threshold relative to the tissue centroid (colored dots; normalized by the maximal distance), for V=0 (left of red line) and V=40V (right of red line) for the same experiment as in Fig. 1a. The same threshold is utilized for both conditions. The different colors mark the location of these signals in different quarters as marked schematically on the right (see Methods). Larger markers are used to highlight isolated islands of activity. (b) A time trace of the spatial correlation coefficients at a length-scale range of 0.08<R<0.12R$_{max}$, where R$_{max}$ is defined by the maximal distance between mesh elements at this time point (see main text and Methods). (c) The time-averaged correlation functions for the same experiment, for V=0 (blue) and V=40V (red). The spatial correlations decay to zero at ~0.25 R$_{max}$ for V=40V and on shorter scales for V=0. (d) The distributions (PDF-normalized to area one) of the spatial correlation coefficients at different length-scales for, V=0 (blue) and V=40V (red). Note the extended correlations of the Ca$^{2+}$ signal under the voltage compared to zero voltage.

Patterning however, requires some level of coordination of the $Ca^{2+}$ activity across the tissue, implying that some level of synchronization of its activity is essential for morphogenesis. The question then arises, whether modulations of the long-range electrical communication in the tissue might affect morphogenesis [35, 55, 56]. Figs. 4a,b show time traces of a tissue in the presence of the alcohol molecule *Heptanol* that blocks gap-junctions [57], and following the wash of the drug (separated by the red lines). Fig. 4a shows the projected tissue *area* (black curve) and r the localization of the strongest coarse grained $Ca^{2+}$ signal above an arbitrary threshold marked by different colors in different quarters as in Fig. 3a. The tissue AR and $Ca^{2+}$ fluorescence spatial Vr are shown in Fig. 4b. The tissue's area under *Heptanol* exhibits large changes but does not show the regular sawtooth pattern characterizing an undisturbed regenerating tissue. Under this condition, the AR remains relatively smooth over time while overall the tissue retains its spheroidal shape and does not exhibit persistent elongations nor a transition in morphology to a cylindrical shape. In this experiment, the tissue does not regenerate under *Heptanol* for the entire duration of the drug application of ~75 hrs (images in Fig. 4c) and regeneration resumes following the wash of the blocker drug only after additional ~41 hrs (images in Fig. 4d). Correspondingly, the $Ca^{2+}$ spatial fluorescence fluctuations signal Vr, shows sparse and highly localized activity under *Heptanol* (Fig. 4a-left of red line). This severe perturbation is reversible. Resumption of the regeneration process is marked by large aspect-ratio fluctuations and a morphological transition to a persistent elongated cylinder characterizing the *Hydra* body-plan, eventually resulting in the completion of the regeneration process (Fig. 4d; Supplemental Movies 5,6 and Fig. S6). More intensive spiking activity resumes after wash of the blocker drug with only moderate localization similar to the situation under zero voltage (Fig.4a-right of red line). The strong localization and low activity of the $Ca^{2+}$ signal under *Heptanol* demonstrates that indeed blocking the gap-junctions disables long-range communication across the tissue which in turn halts morphogenesis. The *Heptanol* manipulation allows us a different observational angle on the tissue's dynamic under conditions that cause pausing of morphogenesis and the inhibition of the morphological transition characterizing regeneration. In that respect they complement the voltage manipulation since now the system cannot undergo the morphological transition due to a weaker $Ca^{2+}$ activity and its strong spatial localization, rather than enhanced activity and extended synchronization under voltage. In both cases, under voltage and under *Heptanol*, the external manipulation prevents proper patterning of the $Ca^{2+}$ activity and inhibits the morphology patterning.

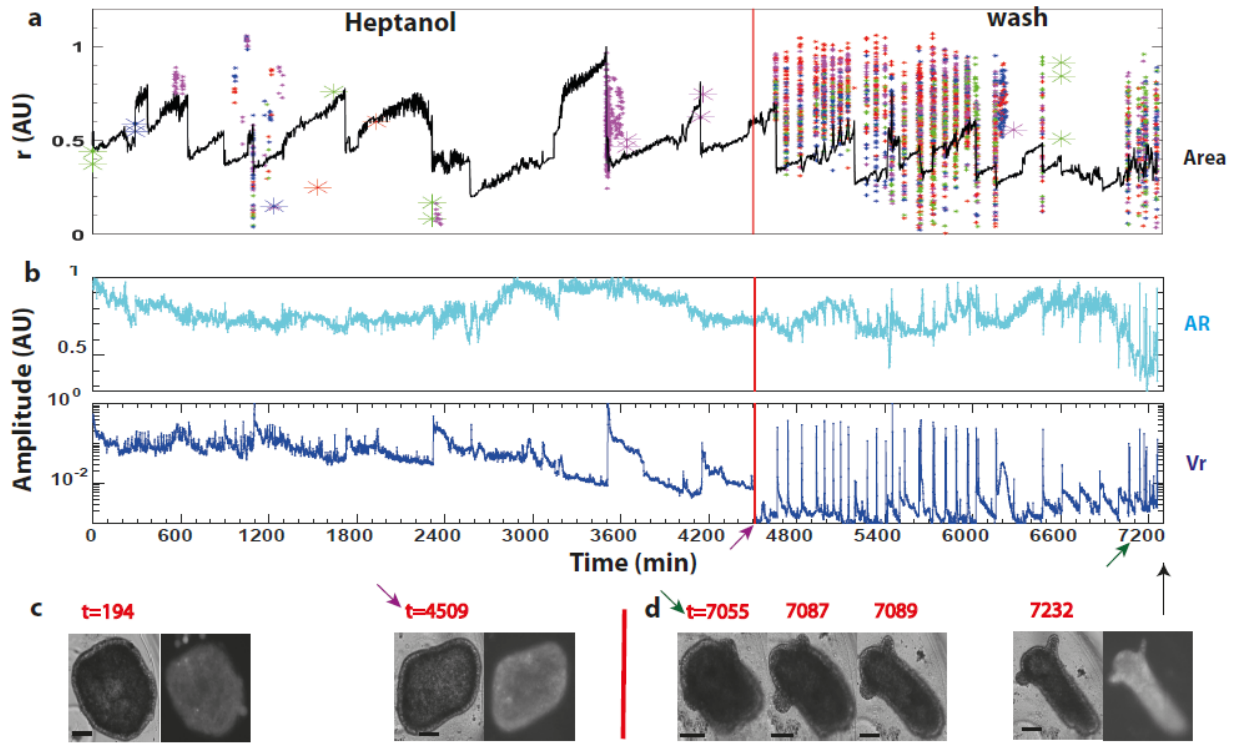

**Fig. 4: Blocking the gap-junctions long-range communication.** (a) Time traces of the projected *area* (black curve; normalized by the max over the trace) and r, the location of the coarse-grained $Ca^{2+}$ signals above threshold relative to the tissue centroid (colored dots; normalized by the maximal distance) for a tissue under *Heptanol* (left of red line) and after wash of the drug (right of red line). The same threshold is utilized for both conditions. The different colors determine different quarters of the tissue as in Fig. 3a. Note the dilute appearance of significant signals and their strong localization under the blocker, in comparison to their abundance and more extended locations after its wash. Larger markers are used to highlight isolated islands of activity. (b) Time traces of the AR and the $Ca^{2+}$ spatial Vr (normalized by the max over the trace; note the y-axis lo scale) under *Heptanol* (left of red line) and following wash of the drug (right of red line). (c) Pairs of BF and fluorescence images of the tissue at the beginning of the trace under *Heptanol* (left) and before washing the drug (right) at the red line marked on the traces (purple arrow). Note the spheroidal shape of the tissue that does not regenerate under this condition. (d) BF images of the tissue showing a morphological transition from a spheroid to a persistent elongated cylinder after washing the *Heptanol* (left; green arrow) and a pair of BF and fluoresence images at the end of the trace showing the fully regenerated *Hydra* (black arrow). Bars: 100μm.

Till now we showed the spatio-temporal characteristics of the $Ca^{2+}$ dynamics and their relation to the morphology patterning. Our conclusion is that the $Ca^{2+}$ spatial fluctuations, under certain spatial distributions and level of correlations, serve as the relevant field associated with morphology patterning and affecting the morphological transition in regeneration. To better understand the nature of these spatial fluctuations and gain a comprehensive view of them, we next discuss their global statistical characteristics by compiling data from multiple tissue samples across different conditions. Fig. 5 shows the spatial fluorescence Vr (Fig. 5a-left) and the spatial Std/mean (Fig. 5b-left) distributions (PDFs), computed for accumulated statistics over different tissue samples, for different experimental conditions, with and without an external field and *Heptanol*. Different conditions lead to significant changes in the distributions of the $Ca^{2+}$ spatial fluctuation signals; applications of an external voltage leads to the most extended tails, while the *Heptanol* application leads to much narrower distributions. This observation is compatible with the results shown above. The global statistical view of the distributions indicates an important character of the $Ca^{2+}$ spatial fluctuations and the underlying constrains over them. While the distributions of the spatial Vr and Std/mean are quite different and show variability across conditions, they do show common global characteristics; all distributions are strictly *non-Gaussian*, skewed and exhibiting extended exponential tails. Different tissue samples under the same condition also exhibit large variability in their $Ca^{2+}$ fluctuation distributions, but they are all showing extended exponential tails (Fig. S7). To search for universal characteristics shared by the distributions, we regard each time trace as a statistical ensemble and normalize the different distributions of the spatial Vr and spatial Std/mean signals by subtracting their corresponding *ensemble-averages* and dividing by their *ensemble standard-deviations* (Figs. 5a,b-right). Remarkably, all the spatial Std/mean distributions for all conditions collapse into a similar shape distribution under this normalization. By contrast, the spatial Vr distributions do not collapse under the same normalization. Similarly, none of the other $Ca^{2+}$ observables, the spatial mean and the total fluorescence signals, collapse into a single shape distribution (Fig. S8).

Interestingly, the best fit to the ensemble-normalized spatial Std/mean statistics is the *Frechet* distribution, belonging to the family of extreme value distributions [58]. This does not necessarily imply that the actual mechanism determining the normalized fluctuations is due to extreme values. Rather, it might reflect underlying correlations and suggests that the spatial fluctuations in the $Ca^{2+}$ activity are constrained in a certain universal way. The fact that the spatial Std/mean distributions, the level of normalized fluctuations, are the only ones that show a collapse to a particular single shape under the ensemble normalization, means that this measure of the $Ca^{2+}$ activity fluctuations is indeed a unique universal observable constrained in the tissue. This is presumably also related to the relations between the first two moments, variance and mean, of the spatial $Ca^{2+}$ activity. Indeed, we note that the spatial variance and mean are correlated and roughly follow a power-law relation: Vr~(mean)$^{2.5}$, with a similar power under all conditions (Figs. 5a,b-insets). This implies that the spatial Std/mean is itself a non-trivial function of the spatial mean signal. It again suggests that the level of $Ca^{2+}$ activity and the synchronization of this signal across the tissue are connected; *a spike of activity emerges in the tissue by an increase in the spatial level of fluctuations rather than by a uniform enhancement of the global activity*. The relaxation of a spike in the $Ca^{2+}$ activity is similarly realized by its decay into sparse localized islands of activity rather than in a global uniform manner. This picture of the $Ca^{2+}$ fluctuations is compatible with the analysis presented above (Figs. 3,4 and Figs. S3,S5). This behavior is not typical of other types of gap-junction coupled electrically excitable systems, which usually recruit resources to fire a synchronized extended excitation and then relax by a global uniform decay of the activity [59].

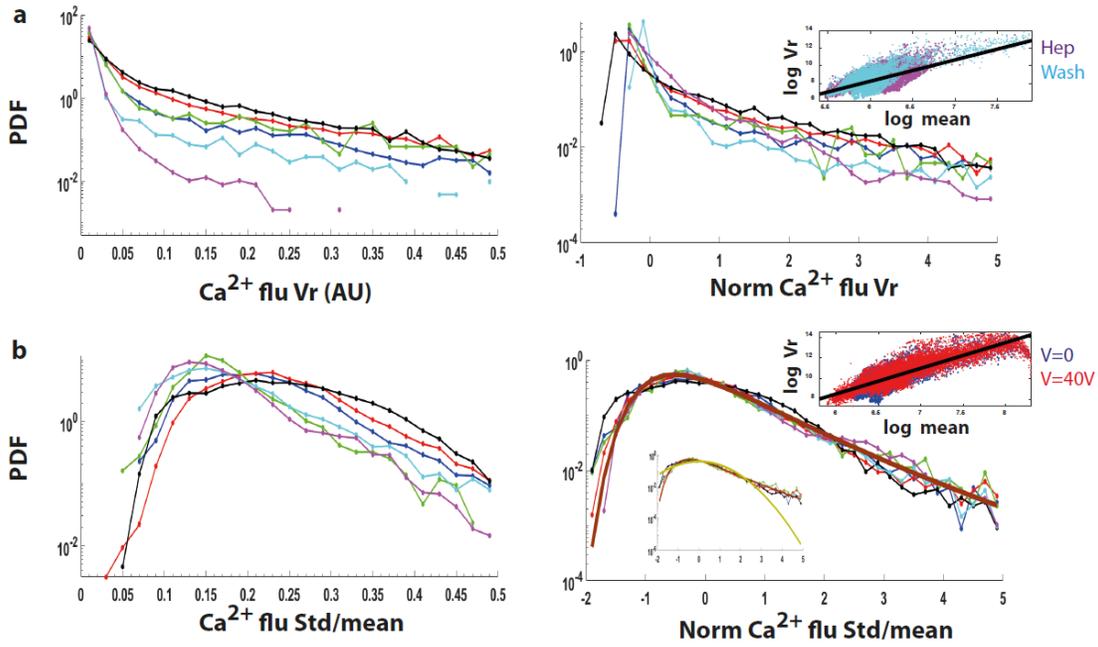

**Fig. 5: Statistics of the Ca²⁺ fluorescence spatial fluctuations.** (a-left) The spatial Vr distributions (PDF-normalized to area one) under different conditions: V=0 (blue), V=40V (red), V=0 after a period under voltage (green), V~50V after a second voltage switch (black), under *Heptanol* (magenta) and following the wash of the drug (cyan). The Vr signals are normalized by their maximum values along each trace. Each histogram is computed from more than 10,000 measurement points (except the switch to V=0 following a voltage period which contains 2,200 points), accumulated from different tissue samples (6-10 samples in 2-3 separate experiments). (a-right) The Vr distributions (PDFs) following normalizations. The signals are normalized by regarding the time-traces as statistical ensembles, subtracting the ensemble-average and dividing by the ensemble standard-deviation of Vr to get standard signals (Methods). Inset: Scatter plot of log spatial Vr versus log spatial mean under *Heptanol* (magenta) and after wash (cyan). Each data point corresponds to a single measurement along the time trace. (b-left) The spatial Std/mean distributions (PDFs) for the same measurements as in (a). (b-right) The spatial Std/mean distributions normalized to standard values as in (a-right). Note the collapse of the distributions into a similar shape. The brown curve is the best fit to that shape, represented by the generalized extreme value statistics PDF:

$$(1/\sigma)\exp\left[-\left(1+k\frac{(x-\mu)}{\sigma}\right)^{-1/k}\right]\left(1+k\frac{(x-\mu)}{\sigma}\right)^{-1-1/k}$$

with parameters, k=0.03, $\sigma = 0.73$ and $\mu = -0.44123$, corresponding to the *Frechet* distribution. Lower inset: a fit of the same data to a Gaussian distribution (yellow). All distributions are strictly non-Gaussian, skewed and exhibiting long exponential tails. Top inset: Scatter plot of log spatial Vr versus log spatial mean for V=0 (blue) and V=40V (red). Each data point corresponds to a single measurement along the time trace. The scatter plots in both insets show that the spatial Vr is approximately a power-law of the spatial mean with the best fit power of 2.5 (black lines; $R^2$~0.5) for all conditions.

**Concluding comments**

In this work we demonstrate a methodology to overcome some of the limitations towards an experimental investigation of the physics of morphogenesis as a pattern formation process. Taking advantage of the flexibility to modulate morphogenesis in whole-body *Hydra* regeneration from a tissue segment, we utilize external electric fields as effective controls allowing us to study the dynamics at the onset of morphogenesis for extended periods. The resulting statistical characterization shows that the $Ca^{2+}$ spatial fluctuations are strongly correlated with the morphology dynamics and possibly play a crucial role in the tissue patterning.

Tissue regeneration involves considerable morphology dynamics, resulting eventually in a transition from a spheroidal shape to a persistent elongated cylindrical shape, characterizing the body-plan of a mature *Hydra*. Such a morphological transition requires moderate $Ca^{2+}$ activity that is weakly correlated across the tissue. Application of an external voltage pauses regeneration at the onset of morphogenesis by increasing significantly the overall $Ca^{2+}$ activity, characterized by fast temporal fluctuations as well as more extended spatial correlations. This change in behavior caused by the external voltage is reversible, allowing us to explore the critical change of morphology from both sides of the transition. The time required for the tissue to recover and resume regeneration after switching the voltage to zero is much shorter than the normal regeneration time for undisturbed tissues [11]. The short recovery time suggests that indeed, in the experiments presented here, the observed $Ca^{2+}$ and morphology fluctuations reflect the dynamics at the onset of a morphological transition in the regeneration process. The long-range electrical communication is essential for morphology patterning in regeneration. Blocking the gap-junctions communication allows a complementary view of the morphological transition by reversibly halting regeneration, now associated with a severe reduction in the overall $Ca^{2+}$ activity, characterized by spatially strongly localized sporadic spikes in the tissue.

Grossly, two types of forces determine the mechanical activity underlying the tissue morphology in *Hydra* regeneration: hydrostatic pressure modulated by the osmotic pressure gradients across the epithelial shell and active internal contractile forces generated by the supracellular actin fibers and their associated myosin in the epithelial muscle systems. While the osmotic pressure gradients are thought to work more-or-less isotopically [24], the actomyosin system has a definite axis-alignment determined by the supracellular actin-fiber organization [10, 11]. The active actomyosin force generation is thought to depend directly on the local $Ca^{2+}$ activity [60]. A plausible picture is that the spatio-temporal $Ca^{2+}$ fluctuations modulate these forces, leading to a subtle interplay in the balance of the overall mechanical stress distributions which affects the tissue morphology [61]. However, beyond its direct effect on the actomyosin force generation, $Ca^{2+}$ might also affect multiple other processes, which in turn can lead to morphology patterning by modifying the underlying mechanics of the tissue [62, 63]. Both the osmotic pressure gradients and the actomyosin forces affect and are affected by water flows and ionic currents through the epithelium tissues (both ectoderm and endoderm layers) [64, 65]. $Ca^{2+}$ plays multiple roles in these processes, but unfortunately our concrete knowledge of those and the underlying mechanisms of $Ca^{2+}$ activity in *Hydra* is still lacking. The best analogy of the *Hydra* tissue is to smooth muscles [42]; this analogy however, is rather limited because of the specificity of the *Hydra* physiology and the fact that in the case of a tissue segment, the muscle is undergoing a regeneration process. In light of the strong correlations of the $Ca^{2+}$ activity with the morphology dynamics shown here and the previously demonstrated strong associations between its activity and the underlying electrical activity of the epithelial tissue, $Ca^{2+}$ in *Hydra* regeneration indeed works as a mediator between the electrical and the mechanical processes. In that respect the $Ca^{2+}$ spatial fluctuations serve as an organizing field of the electrical-mechanical sector of morphogenesis. We repeatedly find a characteristic time scale of approximately 100 min in the fluctuation dynamics whose origin is still unknown. In principle it can originate from electrical or mechanical processes or from their couplings. High-frequency $Ca^{2+}$ fluctuations emerging under an external voltage are presumably averaged out over the characteristic mechanical relaxation times, preventing an effective patterning of the mechanical stresses.

Our experiments demonstrate that the $Ca^{2+}$ spatial fluctuations are strictly *non-Gaussian* under all conditions, suggesting that they do not simply arise from a sum of independent random processes. The actual levels of $Ca^{2+}$ fluctuations, as reflected in their statistical distributions, are highly sensitive to the external and internal conditions of the tissue. However, the shape distribution of the ensemble-normalized Std/mean fluctuations is universally constrained by some yet unknown mechanism, dictating statistics resembling the universal extreme value distributions. Interestingly, such shape distributions are also found in various different physical systems as well as in living ones, indicating underlying correlations in the signals and reflecting their non-stationary nature, but a theoretical understanding of their universal appearance in such systems is still lacking [66-69].

The role of $Ca^{2+}$ fluctuations in determining the morphology of the tissue and its dynamics suggests a picture of morphogenesis as a *dynamical phase transition*, driven by the underlying field fluctuations [70, 71]. The methodology of controlling the dynamics at the onset of transition points should be further developed to enable inquiry of other organizing fields integrating the biochemical, mechanical and electrical processes underlying morphogenesis. It could possibly also be extended to other organisms which might be more constrained than *Hydra* but still allowing considerable manipulations.

**Acknowledgements**

I thank Kinneret Keren for helpful discussions and comments on the manuscript. I thank Yitzhak Rabin, Omri Barak, Naama Brenner, Shimon Marom and Yariv Kafri for comments on the manuscript. I thank our lab members: Liora Garion, Yonit Maroudas-Sacks and Lital Shani-Zerbib, for technical help. Special thanks to Gdalyahu Ben-Yoseph for superb technical help in designing and constructing the experimental setup. I thank Anatoly Meller for constructing the electrical control system.
This work was supported by a grant from the Israel Science Foundation (grant No. 228/17).

## Methods

### Hydra strains, culture and sample preparation

Experiments are carried out with a transgenic strain of *Hydra Vulgaris* (*AEP*) carrying a GCaMP6s probe for $Ca^{2+}$, generated by us in the Kiel center [72] using a modified version of the pHyVec1 plasmid which replaces the GFP sequence with a GCaMP6s sequence that was codon-optimized for Hydra (HyGCaMP6s was a gift from R. Yuste lab (Addgene plasmid # 102558 ; http://n2t.net/addgene:102558 ; RRID:Addgene_102558) [53]). See ref [13] for more details. Animals are cultivated in a *Hydra* culture medium (1mM NaHCO3, 1mM CaCl2, 0.1mM MgCl2, 0.1mM KCl, 1mM Tris-HCl pH 7.7) at 18°C. The animals are fed every other day with live *Artemia nauplii* and washed after ~4 hours. Experiments are initiated ~24 hours after feeding. Tissue segments are excised from the middle of a mature *Hydra* using a scalpel equipped with a #15 blade. To obtain fragments, a ring is cut into ~4 parts by additional longitudinal cuts. Fragments are incubated in a dish for ~3 hrs to allow their folding into spheroids prior to transferring them into the experimental sample holder. Regeneration is defined as the appearance of tentacles.

### Sample holder

Spheroid tissues are placed within wells of ~1.3 mm diameter made in a strip of 2% agarose gel (Sigma) to keep the regenerating *Hydra* in place during time lapse imaging. The tissue spheroid, typically of a few hundred microns in size, is free to move within the well. The agarose strip containing 14 wells, is fixed on a transparent plexiglass bar of 1 mm height, anchored to a homemade sample holder. Each well has two platinum mesh electrodes (Platinum gauze 52 mesh, 0.1 mm dia. Wire; Alfa Aesar, Lancashire UK) fixed at its two sides at a distance of 4 mm between them, on two ceramic filled polyether ether ketone (CMF Peek) holders. The separated electrode pairs for each well allows flexibility in the voltage application. A channel on each side separates the sample wells from the electrodes allowing for medium flow. Each electrode pair covers the entire length of the well and their height ensures full coverage of the tissue samples. A peristaltic pump (IPC, Ismatec, Futtererstr, Germany) flows the medium continuously from an external reservoir (replaced at least once every 24 hrs) at a rate of 170 ml/hr into each of the channels between the electrodes and the samples. The medium covers the entire preparation and the volume in the bath is kept fixed throughout the experiments, by pumping medium out from 4 holes which determine the height of the fluid. The continuous medium flow ensures stable environmental conditions and the fixed volume of medium in the bath ensures constant conductivity between the electrodes (measured by the stable current between the electrodes when external voltage is applied). All the experiments are done at room temperature.

### AC generator, multiplexer and the voltage protocol

A computer controlled function generator (PM5138A, Fluke, Everett, WA USA) connected to a voltage amplifier (F20A, FLC Electronics) is used to set the AC voltage between the electrodes. A homemade software (utilizing Labview) controls the voltage between the electrodes and monitors the generated current in the system. Monitoring the current allows us to verify the stabilities of the system's conductivity and the environmental conditions. A switch unit (Keysight 34972A) is controlled by the software to multiplex the applied voltage to the different electrodes in the system. The experiments in this work are done with an AC field of 2 kHz with a similar protocol for all experiments. The voltage change is done gradually at a constant rate to avoid damage to the tissue sample, until it reaches a new set point and held constant thereafter. The AC voltage is applied alternately with 15 sec cycles to the electrodes. Test experiments with continuous application of the voltage give similar results (see also ref [13]).

**Microscopy**

Time lapse bright-field and fluorescence images are taken by a Zeiss Axio-observer microscope (Zeiss, Oberkochen Germany) with a 5× air objective (NA=0.25) with a 1.6× optovar and acquired by a CCD camera (Zyla 5.5 sCMOS, Andor, Belfast, Northern Ireland). The sample holder is placed on a movable stage (Marzhauser, Germany) and the entire microscopy system is operated by Micromanager, recording images at 1 min intervals. The fluorescence recordings at 1 min resolution is chosen on the one hand to allow long experiment while preventing tissue damage throughout the experiments and on the other hand to enable recordings from multiple tissue samples.

**Data Analysis**

*Shape and fluorescence analysis*

For the analysis, images are reduced to 696x520 pixels using ImageJ. Masks depicting the projected tissue shape are determined for a time-lapse movie using the bright-field (BF) images by a segmentation algorithm described in [73] and a custom code written in Matlab. Shape analysis of regenerating Hydra is done by representing the projected shape of the tissue by polygonal outlines using the Celltool package developed by Zach Pincus [74]. The polygons derived from the masks provide a series of (x,y) points corresponding to the tissue's boundary. Each series is resampled to 200 points (or 30 points for mesh generation, see below), which are evenly spaced along the boundary. The polygons generated by this analysis are used to compute the tissue projected area, its aspect-ratio and the tissue centroid by Matlab in the following way. The morphological observables of the tissue, its projected area and aspect ratio are computed using the geom2d package in Matlab[75], with the functions *polygonArea* and *polygonCentroid,* respectively. The tissue centroid is the center of mass of the boundary polygon. The tissue aspect ratio (AR) is computed by finding the best-fit ellipsoid using *polygonSecondAreaMoments* which computes the second-order moment of the boundary polygon and defining the AR as the ratio between the short to long axes of this ellipsoid (i.e., as the values get lower the AR gets larger; AR=1 is a sphere).

The fluorescence analysis is done on images reduced to the same size as the bright-field ones (696x520 pixels). The fluorescence observables are computed by using the mask generated by the shape analysis above to extract the total fluorescence, the mean fluorescence, the spatial variance and spatial standard-deviation over mean using the Matlab standard functions. The fluorescence signals are extracted directly from the Tiff images using Matlab. Temporal cross-correlation and autocorrelation functions are computed from the fluorescence signal after subtracting its spatial mean value (averages over the tissue at each time point) using the Matlab function *XCORR* with unbiased normalization and normalized to 1 at a zero lag. The correlations of the surrogate data marking the level of computational noise is computed from a random permutation of the time trace. To amplify the largest peaks of activity in Fig. 2, we smoothed the traces using the Matlab function *SMOOTHDATA* with a Gaussian-weighted moving average over a window of 10 data points (10 min).

*Mesh generation, analysis of spatial correlations and localization*

We use a Matlab code for generating a triangular mesh over the tissue in order to coarse-grain the fluorescence signal (based on the algorithm in [54] with modifications). The original boundary polygon is first dilated slightly parallel to itself to avoid edge effects and resampled such that its vertices are equally distributed. Since the tissue is a deformable object and free to rotate in 3D, its 2D projected area is not fixed, so the precise number of mesh triangles can change between time points. At each time point, the number of mesh triangles is large enough (usually around 100 or more) to gain enough statistics as well as a reasonable spatial resolution of the coarse-grained fluorescence signals. For each triangular mesh element we compute its mean fluorescence and centroid defined as the center of mass of the triangular polygon.

The spatial correlation functions are computed using these coarse-grained fluorescence signals. At each time point, we first compute a distance matrix between all centroid pairs of the triangles in the mesh and normalize it by the maximal distance $R_{max}$ (note that the distances are now defined by the centroids of the mesh elements). We next compute a covariance matrix between all triangle pairs using a normalized signal for each triangle; subtracting from it the overall tissue mean fluorescence and dividing by its standard-deviation. The normalized distance is divided into bins and the normalized covariance coefficients are averaged within a bin to get an estimate of the correlation coefficient at this distance range. Different partitions into bins do not significantly change the results.

To estimate the level of localization (Figs. 3a,4a) we define an arbitrary threshold and compute the radii r of all triangles' centroids in the mesh, relative to the tissue's centroid, and their x,y coordinates, containing a coarse-grained fluorescence level above this threshold. The location radii are normalized by the tissue length-scale computed as $\sqrt{area/\pi}$. Finally, the tissue is arbitrarily divided in the lab frame into 4 quarters and the location r is plotted for all triangles containing fluorescence signals above threshold, marking the assigned quarters (defined by the triangular mesh x-y centroid coordinates) by different colors. We keep the same quarters' definition in the lab frame over time. The threshold value is varied over a reasonable range to check that the observed results are qualitatively robust and do not depend sensitively on the chosen threshold. Although the tissue is free to rotate and deform, the projected image in 2D is usually stable for long periods, so the defined quarters do not change frequently in the tissue frame over time.

*Fluorescence observables distributions and normalization*

The fluorescence observables distributions are computed in Matlab using the function *HISTCOUNTS* with normalization as a *PDF*. The normalized distributions are computed by treating the time trace as a statistical ensemble, subtracting from the relevant signal its ensemble average and dividing by its ensemble standard-deviation to get standard signals.

Supplementary Material

Movies

**Movie 1: A tissue spheroid at zero voltage.** Pairs of bright-field (BF) and fluorescence images of a tissue spheroid under zero voltage (traces of Fig. 1b, left of the red line).

**Movie 2: A tissue spheroid under 40V.** Pairs of bright-field (BF) and fluorescence images of a tissue spheroid under V=40V (traces of Fig. 1b, right of the red line).

**Movie 3: A tissue spheroid following reset of the applied voltage.** Pairs of bright-field (BF) and fluorescence images of the same tissue spheroid as in Movies 1,2, under zero voltage following a session under V=40V (traces of Fig. 1d, left of the red line).

**Movie 4: A mature *Hydra* under re-application of the external voltage.** Pairs of bright-field (BF) and fluorescence images of a mature *Hydra* regenerated following the reset of the external voltage (Movie 3), under re-application of the voltage at V=50V (traces of Fig. 1d, right of the red line).

**Movie 5: A tissue spheroid under *Heptnol*.** Pairs of bright-field (BF) and fluorescence images of a tissue spheroid under the gap-junction blocker *Heptanol* (traces of Figs. 4a,b, left of the red line).

**Movie 6: A tissue spheroid following the wash of *Heptnol*.** Pairs of bright-field (BF) and fluorescence images of the tissue spheroid of Movie 5, following the wash of the gap-junction blocker *Heptanol* (traces of Figs. 4a,b, right of the red line).

**Supplementary Figures**

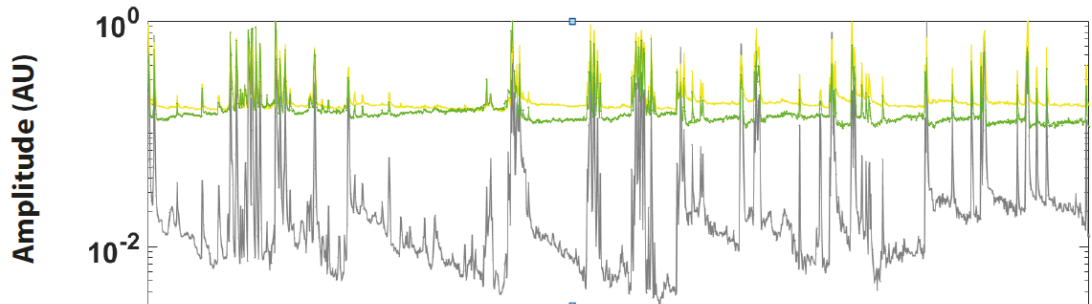

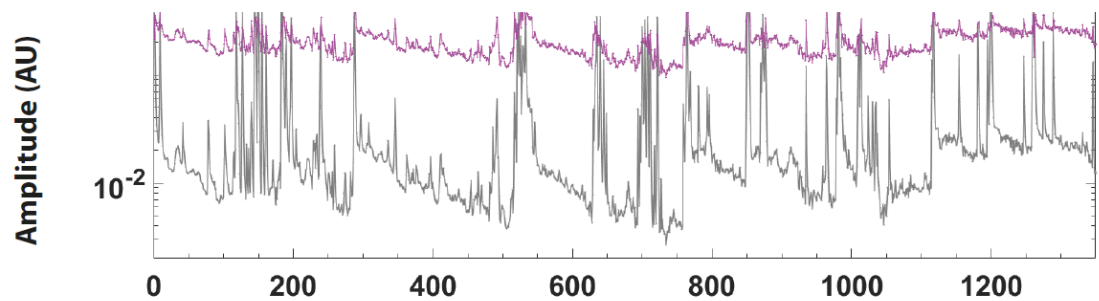

**Fig. S1: Comparison of different Ca$^{2+}$ fluorescence observables.** (a) Total (green) and mean (yellow) fluorescence traces compared with the spatial fluorescence variance, Vr (gray) for the same experiment as in Fig.1b in the main text for V=0. While the main peaks of activity are similarly observed for all the different measures of the Ca$^{2+}$ fluorescence signal, the Vr shows much richer dynamics which cannot be observed in the total or mean fluorescence signals. This is partially due to the fact that the former observables are less sensitive to local changes in the fluorescence signal. The differences between the different observables also highlight the characteristics of the ca$^{2+}$ activity which is relatively localized and highly non-uniform across the tissue. (b) The spatial standard-deviation over mean (Std/mean; magenta) dynamics of the Ca$^{2+}$ fluorescence compared with the spatial Vr (gray) for the same trace as in (a). These two observables show high similarity in their dynamics, with important differences between their statistical characteristics (see the discussion in the main text Fig. 4). Note the y-axis log scale.

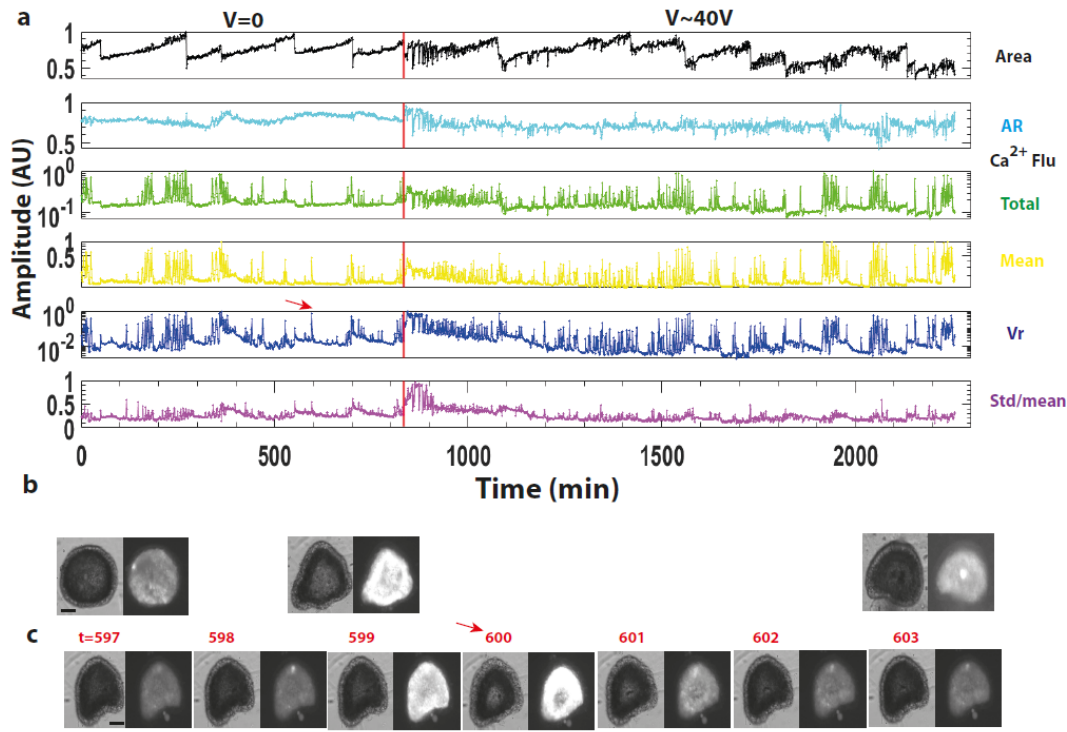

**Fig. S2a: Morphology and $Ca^{2+}$ fluctuation dynamics.** (a) An example of the time traces measured at each time point at 1 min resolution of a tissue sample from a different experiment than the one depicted in Fig. 1 in the main text. Traces from the top: projected *area* of the tissue (normalized by the max over the trace), *aspect-ratio* (AR; short axis/long axis of a corresponding ellipsoid); $Ca^{2+}$ fluorescence observables: *total, mean* and spatial *variance* (Vr) of the florescence signal across the tissue (all signals normalized by their max values over the trace; note the y-axis log scale), and the spatial *standard-deviation over mean* (Std/mean) of the fluorescence signal. The traces to the left of the red line are at zero voltage while those to the right follow the switch of the voltage to ~40V. (b) Pairs of bright-field (BF) and fluorescence images at the beginning of the time traces at zero voltage (left), at the voltage switch marked by the red line (middle) and at the end of the traces under voltage (right). (c): Pairs of BF and fluorescence images around an example peak of the Vr, marked by the red arrow on its trace. Note the localization of the $Ca^{2+}$ fluorescence signal and the non-uniform excitation leading to the emergence of an activity spike and its relaxation. Bars: 100μm.

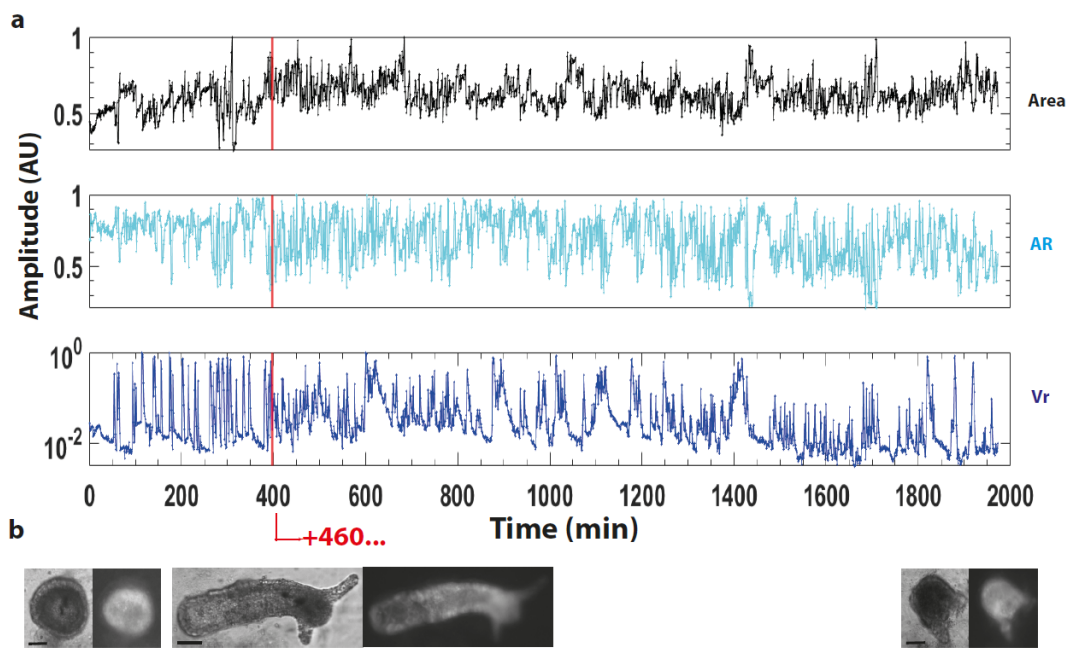

**Fig. S2b: Morphology and Ca²⁺ fluorescence observables at a second voltage switch.** (a) Time traces of the same tissue as in Fig. S2a, following switching-off the external voltage at the end of the V~40V trace (left of red line) and then a second voltage switch to V~50V (at the red line). Upper panel: the tissue projected *area* (normalized by the max over the trace), middle panel: *AR* and lower panel: the Ca²⁺ fluorescence spatial *Vr* (normalized by the max over the trace); note the y-axis log scale. The measurements of the zero voltage trace actually extend to 857 min; for clarity we cut and connect the two traces in Fig. 1c at t=397 min since after that time, the tissue is fully regenerated (reflected in an increase of the tissue area) and the strong movements of the mature animal gets it occasionally out of focus, making the measurements unreliable. (b) Pairs of BF and fluorescence images of the tissue at the beginning of the traces (left), just before the voltage switch at the red line showing a fully regenerated *Hydra* (middle) and at the end of the voltage trace, showing the change in morphology from an elongated cylinder to a spheroid (right). This voltage is not strong enough to completely eliminate the tentacles and causing a full reversal of morphology. Bars: 100μm.

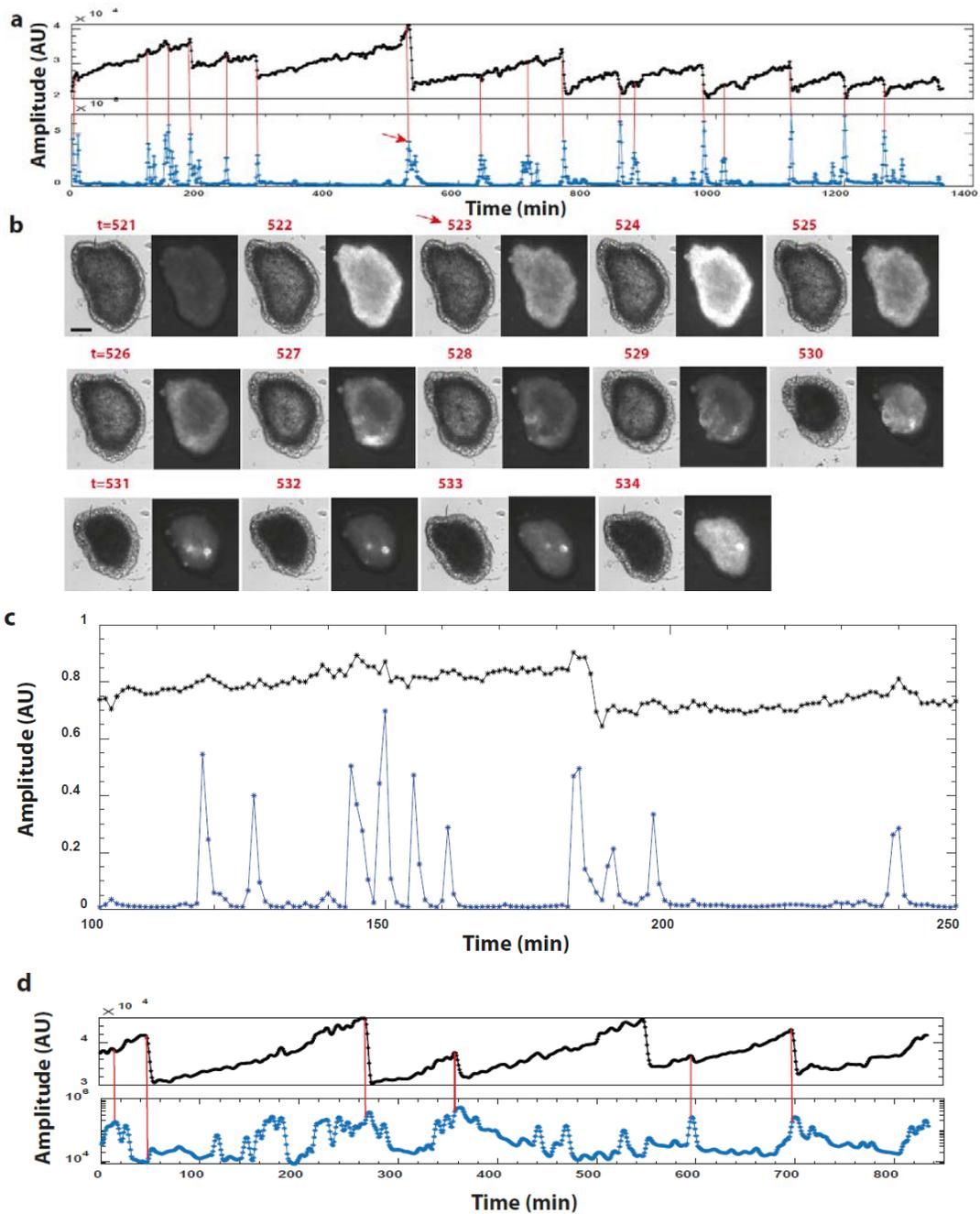

**Fig. S3: Area-Vr associations.** (a) The area (top) and Ca$^{2+}$ fluorescence spatial Vr (bottom), the same as in the top panel of Fig. 2a in the main text. in the main text). The traces are smoothed and similarly, the red lines connect peaks in the Vr with the corresponding time points on the area trace. (b) Pairs of BF and Ca$^{2+}$ fluorescence images showing the emergence of a Vr peak and its decay around the time point marked by a red arrow on the Vr trace. The time markings above the images are in minutes. Note the localization of the Ca$^{2+}$ fluorescence signals. Bars: 100μm. (c) A zoom over a segment of the trace in (a), not smoothed here (normalized by the max over the trace), showing that changes in area are associated with an increase in the fluorescence spatial Vr levels. Note that these changes in the Vr are associated with either increase or decrease of the tissue area and that the increase in Vr precedes the large drop in the tissue's area. (d) Another example of the close associations between the tissue area (top) and the fluorescence Vr (bottom) for the same tissue sample as in Fig. S2a for V=0. The traces are smoothed as in (a).

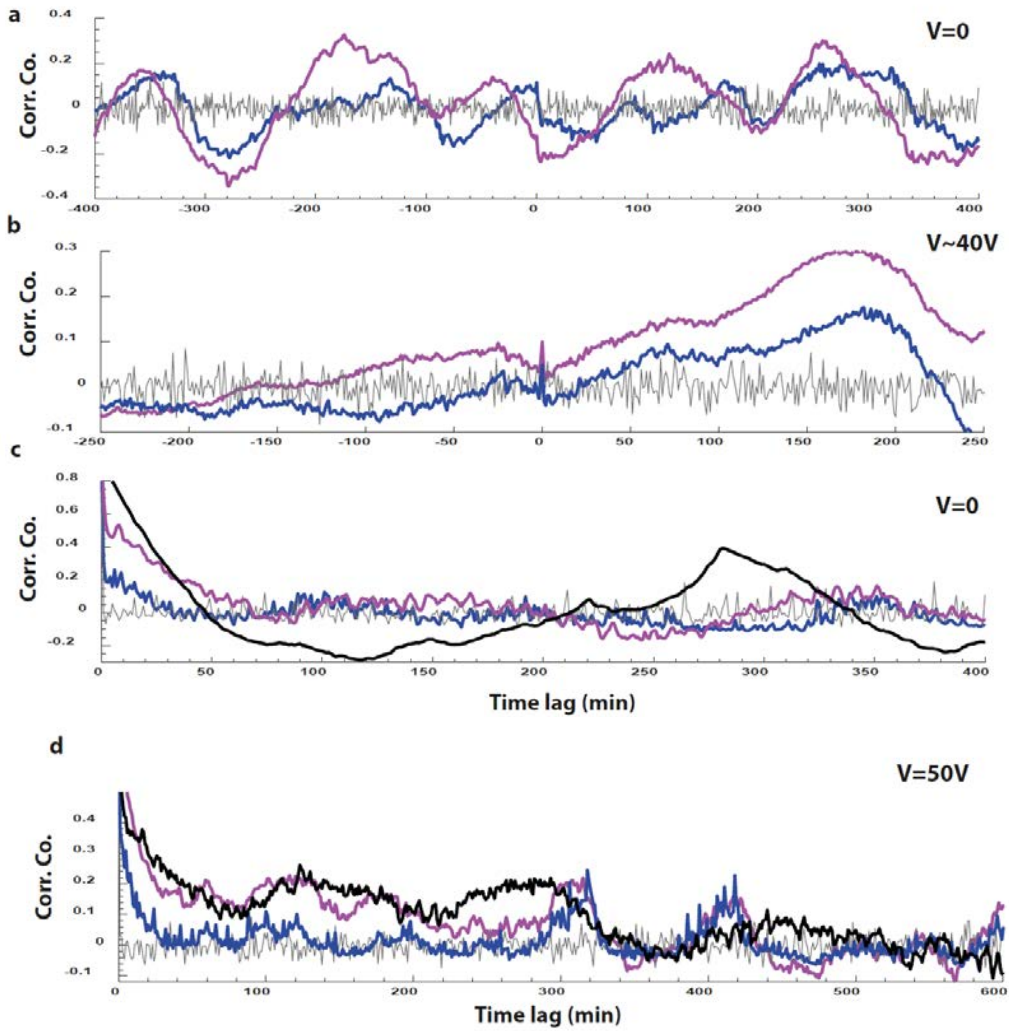

**Fig. S4: Spatial correlations.** Cross-correlation functions between the tissue area and the $Ca^{2+}$ fluorescence spatial Vr (blue) and the spatial Std/mean (magenta) for the same tissue sample as in Fig. S2a, for V=0 (a) and V~40V (b). Note the oscillatory nature of the dynamics at V=0 and the time scale of approximately 100 min. (c) The autocorrelation functions for the same tissue sample at V=0 computed for the area (black), $Ca^{2+}$ fluorescence spatial Vr (blue) and spatial Std/mean (magenta). The analysis is shown for short enough time lags to avoid confusion due to the non-stationarity of the signals. (d) Another example of autocorrelation functions for a different tissue sample at high voltage, for the area (black), the $Ca^{2+}$ fluorescence spatial Vr (blue) and the spatial Std/mean (magenta). The tissue sample went through a round of voltage applications: starting at V=0, switching-on of the voltage to V~40V halting regeneration, switching off the voltage to zero again and then increasing it to V=50V. Note the relaxation time of approximately 100 min for the area and Std/mean and the oscillatory nature of their signals. The gray curves in all autocorrelations and cross-correlations plots is computed from surrogate data, indicating the level of noise in the computations (Methods).

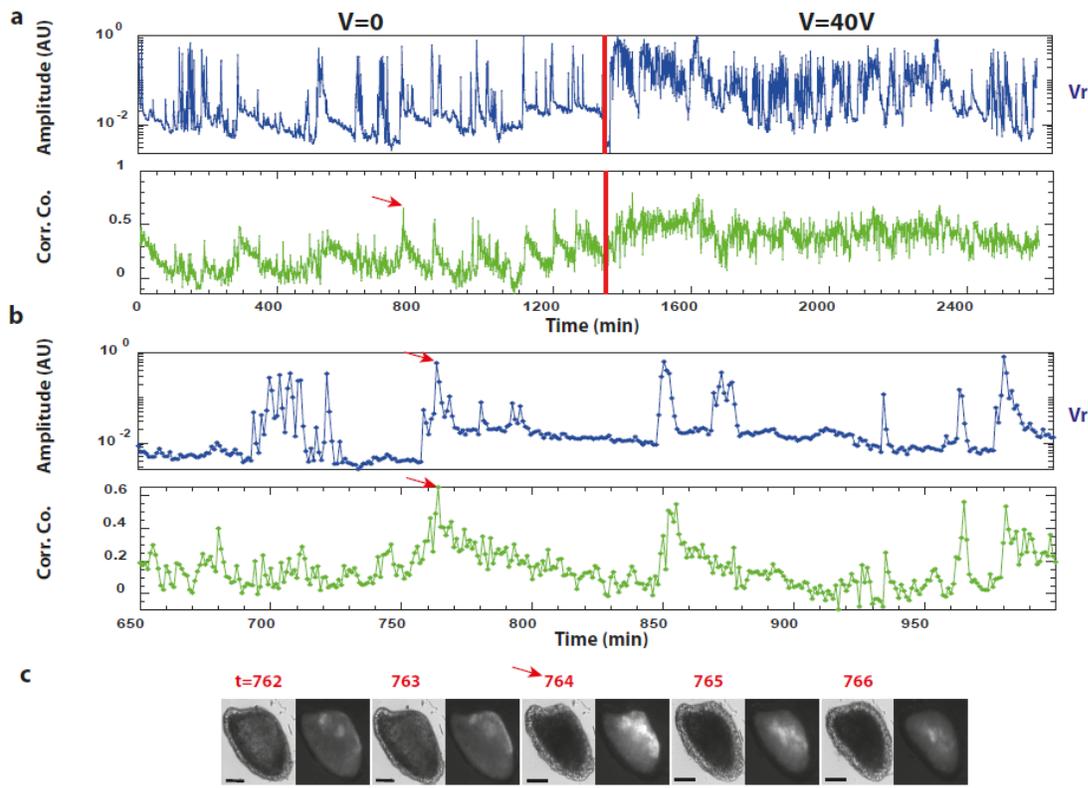

**Fig. S5: Associations of the spatial correlations with the spatial Vr.** (a) Traces of the $Ca^{2+}$ fluorescence spatial Vr (top) and the spatial correlation coefficients (bottom) at a length-scale range $0.08<R<0.12R_{max}$, where $R_{max}$ is defined by the maximal distance between mesh elements at this time point (see main text and Methods), for the same tissue sample as in Fig. 3 in the main text. The Vr signal is normalized by the max over the trace. Note the y-axis log scale for the spatial Vr. The voltage is switched from zero to 40V at the red line. (b) Zoom over a region of the traces in (a) showing a more detailed comparison of the dynamics between the spatial Vr (top) and the spatial correlation coefficients (bottom). Note the y-axis log scale for the spatial Vr. Although there are similarities in the response of these two observables, their dynamics show significant differences. (c) Pairs of BF and fluorescence images showing the tissue sample around a peak in the spatial Vr and spatial correlation coefficients, marked by the red arrows. The time markings above the images are in minutes. Bars: 100μm.

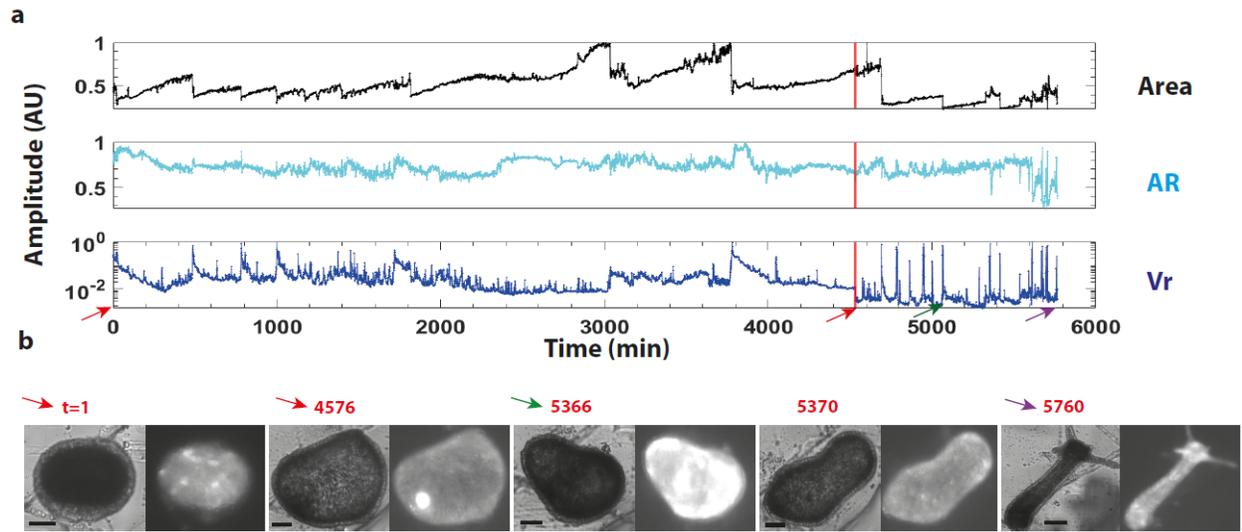

**Fig. S6: Tissue dynamics under *Heptanol*.** (a) Time traces of the tissue projected area (top; normalized by the max over the trace), AR (middle) and the $Ca^{2+}$ fluorescence spatial Vr (bottom; normalized by the max over the trace; y-axis log scale) for another tissue sample than the one shown in Fig. 3 in the main text. The tissue is shown under the gap-junction blocker *Heptanol* (left of red line) and after wash of the drug (right of red line). Note the dilute appearance of significant signals under the blocker drug in comparison to their abundance after its wash. (b) Pairs of BF and fluorescence images of the tissue at the beginning of the trace under *Heptanol* (left; t=1 min, red arrow) and just after washing the drug (t=4576 min, red arrow) at the red line marked on the traces. Note the spheroidal shape of the tissue that does not regenerate under this condition. Images following the wash of *Heptanol* at: t=5366 min, green arrow; t=5370 min and at the end of the trace at t=5760 min, purple arrow. The tissue's morphology is changed following the wash of the drug, showing persistent elongations and eventually a transition from a spheroid to an elongated cylinder characterizing a mature *Hydra*. The time markings above the images are in minutes. Bars: 100μm.

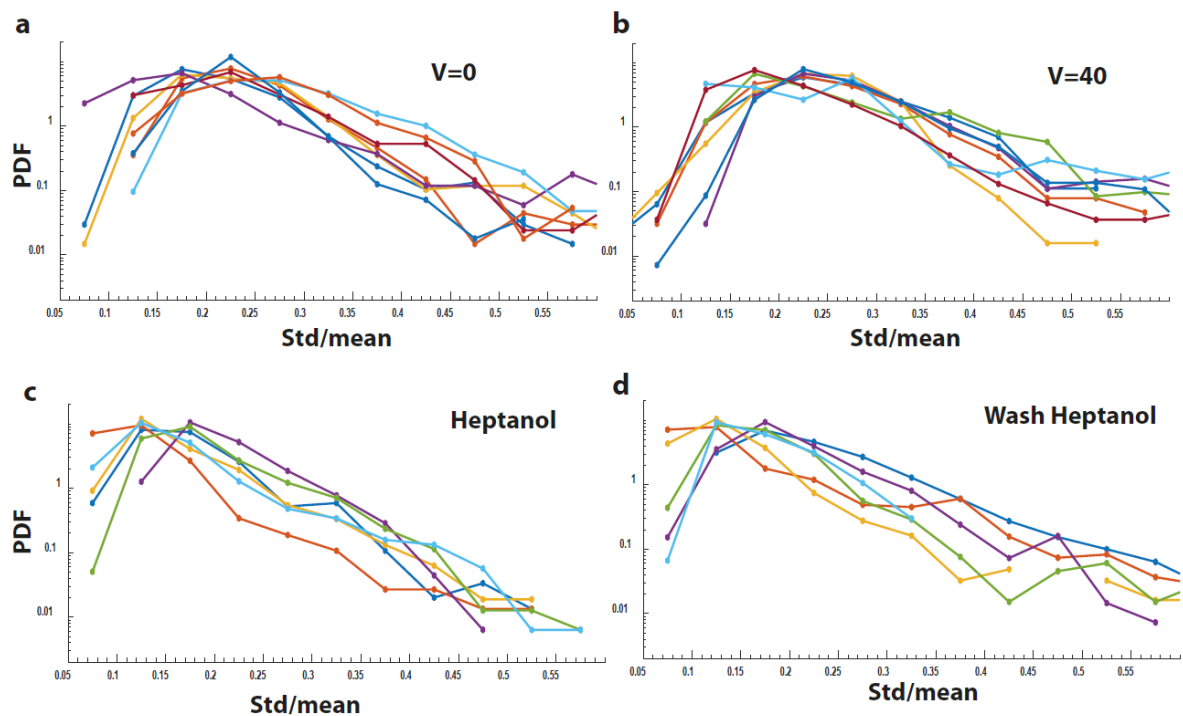

**Fig. S7: Variability of the Ca$^{2+}$ fluorescence spatial Std/mean between tissue samples.** The spatial Std/mean distributions for different tissue samples under the same experimental conditions at (a) V=0, (b) V=40V, (c) under *Heptanol*, and (d) after wash of the drug. All histograms show PDFs normalized to area one. Note the significant variability between different tissue samples under the same condition. Note also that all tissue samples exhibit skewed distributions with extended exponential tails.

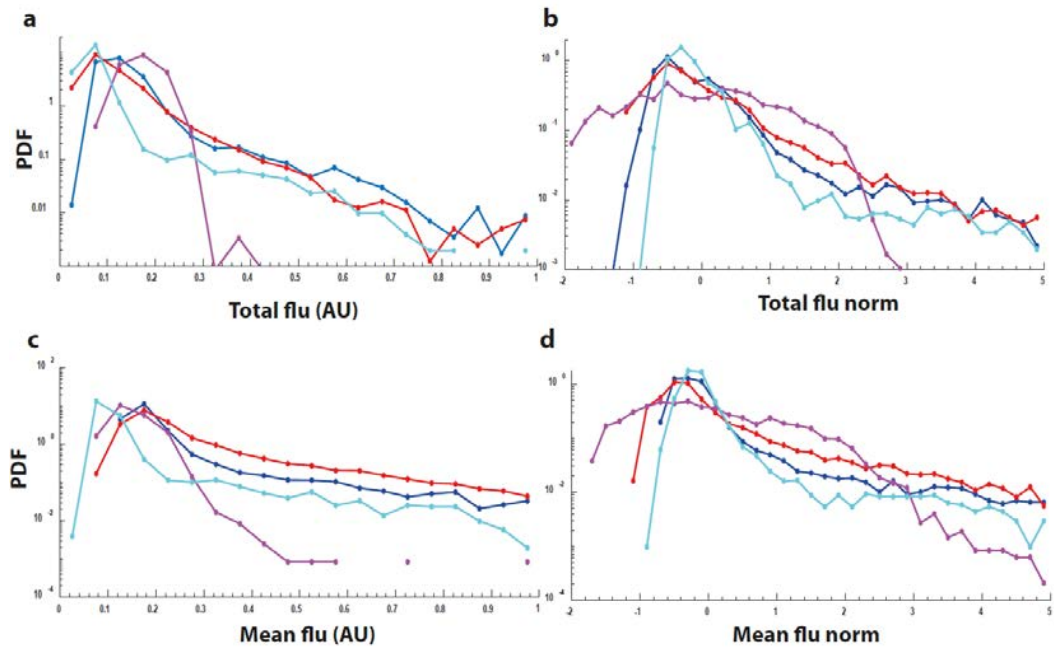

**Fig. S8: Ca²⁺ mean and total fluorescence distributions do not collapse under normalization.** (a) The distributions of total Ca²⁺ fluorescence levels accumulating statistics over samples for different conditions (the same as in Fig. 5 in the main text): V=0 (blue), V=40V (red), under *Heptanol* (magenta) and following the wash of the drug (cyan). (b) The same data as in (a) following normalization. The total fluorescence signals are normalized by regarding the time-traces as statistical ensembles, subtracting the ensemble-average total fluorescence and dividing by its ensemble standard-deviation to get standard signals (as in Fig. 5 in the main text, see Methods). Note that unlike the Std/mean, this observable does not collapse into a universal shape distribution. (c) The spatial mean fluorescence (total fluorescence divided by the projected area) distributions for the same tissue samples and conditions as in (a). (d) Normalizing the mean fluorescence data in (c) similar to the normalization in (b), by subtracting the ensemble-average mean fluorescence and dividing by its ensemble standard-deviation to get standard signals. Note that also the spatial mean fluorescence signals for the different conditions do not collapse to a universal shape distribution.